# THE CORRELATED COLORS OF TRANSNEPTUNIAN BINARIES

S. D. Benecchi[a,1,*], K. S. Noll[a], W. M. Grundy,[b] M. W. Buie,[b,c] D. C. Stephens[d], H. F. Levison[c]

[a] Space Telescope Science Institute, 3700 San Martin Dr., Baltimore, MD 21218

[*] Corresponding Author E-mail address: susank@stsci.edu

[b] Lowell Observatory, 1400 W. Mars Hill Rd., Flagstaff, AZ 86001

[c] Dept. of Space Studies, Southwest Research Institute, 1050 Walnut St. #400, Boulder, CO 80302

[d] Brigham Young University, Dept. of Physics & Astronomy, N145 ESC, Provo, UT 84602

[1] Née S. D. Kern

Pages: 28
Figures: 4
Tables: 4

**Proposed running head:** Colors of transneptunian binaries

**Editorial correspondence to:**

Dr. Susan D. Benecchi
Space Telescope Science Institute
3700 San Martin Dr.
Baltimore MD 21218
Phone: 410-338-4051
Fax: 410-338-5090
E-mail address: susank@stsci.edu

**Abstract**

We report resolved photometry of the primary and secondary components of 23 transneptunian binaries obtained with the Hubble Space Telescope. V-I colors of the components range from 0.7 to 1.5 with a median uncertainty of 0.06 magnitudes. The colors of the primaries and secondaries are correlated with a Spearman rank correlation probability of 99.99991%, 5 sigma for a normal distribution. Fits to the primary vs. secondary colors are identical to within measurement uncertainties. The color range of binaries as a group is indistinguishable from that of the larger population of apparently single transneptunian objects. Whatever mechanism produced the colors of apparently single TNOs acted equally on binary systems. The most likely explanation is that the colors of transneptunian objects and binaries alike are primordial and indicative of their origin in a locally homogeneous, globally heterogeneous protoplanetary disk.

## 1. Introduction

Transneptunian objects (TNOs) have long been known to display a surprisingly diverse range of colors (Jewitt & Luu, 1998). In some cases these colors have been correlated with dynamical properties (e.g. Tegler & Romanishin 2000; Trujillo & Brown 2002; Peixhino et al. 2004; Gulbis et al. 2006; Doressoundiram et al. 2008), although usually at low statistical significance. It has remained uncertain whether the large range of TNO colors is caused by differences in primordial chemical composition or by environmental factors such as the interplay of impact resurfacing and space weathering (Luu & Jewitt, 1996; Thébault & Doressoundiram 2003). Color measurements have been made at both visible and infrared wavelengths (Gil-Hutton & Licandro 2001; Jewitt & Luu 2001; Delsanti et al. 2002 and 2004; Doressoundiram et al. 2002 and 2005; Stephens et al. 2003; McBride et al. 2003; Bauer et al. 2003, Peixinho et al. 2004). The largest compilations of these colors are contained in the MBOSS (Minor Bodies in the Outer Solar System; Hainaut & Delsanti 2002) and HST (Stephens et al. 2007) databases. At the wavelengths relevant to this work (V through J bands), TNOs have been found to be spectrophotometrically indistinct (i.e. they don't display strong spectroscopic features), but they differ from one another in slope (Boehnhardt et al. 2002; Peixinho et al. 2004). Attempts to model the spectrum shortward of 1000nm have generally assumed a variable mixture of red organic material and a spectrally gray material (Brunetto et al. 2008; Grundy 2008c).

Unresolved colors have been obtained for about half of the more than 50 known transneptunian binaries (TNBs), but colors of resolved binary components have been reported for only 8 TNBs (Noll et al. 2008a). For many of the resolved measurements, the uncertainties are large (Noll et al. 2002, 2004ab; Osip 2003) or unstated (Margot et al. 2005). No clear trends are identifiable from this small sample, although in most cases the colors of the components are similar.

In this work we present resolved color photometry of 23 TNBs using images from the Hubble Space Telescope (HST). We report photometry derived from both newly acquired observations and reanalyzed archival data. By using an improved point-spread-function-fitting technique we can achieve smaller uncertainties than previous work. With this new, uniformly reduced data we are able, for the first time, to definitively address the question of the colors of TNBs. As we show, these colors have implications for TNOs as a whole and their origins in the protoplanetary disk.

## 2. Observations

In this work we report color photometry from the Wide Field Planetary Camera 2 (WFPC2) and the Advanced Camera for Surveys High Resolution Camera (HRC) from seven HST programs. The colors of 13 TNBs reported here were measured as part of program 11178 (Grundy et al. 2008b). It used the WFPC2 camera to measure TNB colors during a single epoch (normally spanning two 96-minute HST orbits, occasionally spanning three) using the 'wide V' F606W filter and the 'wide I' F814W filter. The filters are centered at 600.1 nm and 799.6 nm with FWHM of 150.2 nm and 152.2 nm, respectively (Heyer et al. 2004). The observation sequence was two exposures in F606W (260 seconds each) followed by two exposures in F814W (500 seconds each) for the first orbit, and then the same sequence in reverse order for the second orbit. Bracketing the observations by F606W exposures helped to identify possible effects due to short-period lightcurve variability, which we discuss in section 5.4. The four exposures in each filter were dithered using a sub-pixel dither pattern offset by (11.0,5.5), (16.5,16.5) and (5.5,11.0) pixels relative to the first exposure.

We have measured the colors of 19 binaries in program 11178, but in 6 cases the color measurements were obtained when the components were separated by less than one PC pixel (45 milliarcsec). Because of the difficulty in resolving components at such small separations, we have excluded (80806) 2000 $CM_{105}$, (82075) 2000 $YW_{183}$ (119979) 2002 $WC_{19}$, (182933) 2002

$GZ_{31}$, 2001 $FL_{185}$ and 1999 $RT_{214}$ from our analysis of resolved colors, but include them in the comparison of unresolved colors in section 5.1.

HST programs 10508, 9991 and 9746 obtained color measurements of 8 TNBs using the HRC (Margot et al. 2005; Stansberry et al. 2006; Grundy et al. 2007, 2008a). Three additional HST programs 9259,9320 and 9508, obtained WFPC2 observations of a single TNB, 1998 $WW_{31}$, over multiple epochs (Veillet et al. 2002). A summary of each HST program analyzed in this study can be found in Table 1.

**[ Table 1 ]**

For completeness in our analysis we have included the resolved ground based colors of two additional TNBs, (88611) Teharonhiawako/Sawiskera and $2005EO_{304}$. These data were obtained at Las Campanas Observatory from the 6.5-m Clay telescope using the Raymond and Beverly Sackler Magellan Instant Camera (MagIC, Osip et al. 2004) under excellent (<0.5 arcsec) seeing conditions. Analysis of these images is detailed in Kern (2006).

## 3. Data Reduction and analysis

### *3.1. Pipeline Processing*

We processed the data through the standard HST pipeline (Baggett et al. 2002; Pavlovsky et al. 2006). This software performs basic image reduction: it flags static bad pixels, performs A/D correction[1], subtracts the bias, and dark images, and corrects for flat fielding. It also updates the header with the appropriate photometry keywords. The flat-field-calibrated images, subscripted by *c0f* for WFPC2 and *flt* for HRC, were the ones we analyzed.

[1] A/D correction takes the observed charge in each pixel in the CCD and converts it to a digital number with the appropriate gain setting. The gain for WFPC2 is 7.12 and for ACS is 2.216.

### *3.2. Point-spread function fitting*

In simplest terms, our analysis process fits a model point-spread function (PSF) generated by the Tiny Tim program (Krist & Hook 2004, v6.3) to the data to determine the locations of the binary components and to perform photometry. The detailed steps differ slightly between instruments with additional steps being required on ACS images due to its larger geometric distortion. In the following text we describe the steps required for ACS. The WFPC2 analysis is very similar; it is described in full detail in Grundy et al. (2008a).

#### *3.2.1. Object identification and initial values*

We began our PSF-fitting by estimating the center position of each binary component by eye (good to about 0.5 pixels) and by summing the object flux in a 1.5 pixel radius around that position. Calculations were carried out in a sub-image, 20-60 pixels on a side, centered on the primary component. We defined the sub-image to be large enough to encompass both components of the binary and to ensure sufficient sky so that the background can be computed. Bad pixels from cosmic rays and hot pixels within the sub-image were flagged and ignored in further calculations prior to determining the background.

#### *3.2.2. Model images*

We further refined our initial guesses for object positions and fluxes by creating a series of scaled and sub-sampled Tiny Tim PSFs generated for a solar spectral distribution [models 51 (HD 150205, a G5V star) and 57 (HD 154712, a K4V star) from the Bruzual-Persson-Gunn-Stryker Spectra Library; Krist & Hook 2004] around the estimated positions of each component. A new sub-sampled PSF was created for each whole pixel location and sub-pixel shifting was performed before re-sampling the model to match the image pixel scale. Additionally, the PSF was modified to account for changes in the focus due to small thermally-induced changes known as "breathing" and for small motions of the spacecraft known as "jitter" that occur on orbital timescales. The first effect is modeled by the z4 parameter in the input file for Tiny Tim. The second effect is modeled by convolving the model with a smearing Gaussian. The determination of these two effects is iterative after the initial values for the binary components are refined. The focus and smearing values that yield a minimization of the $\chi^2$ residual was defined to be the best focus and smearing value for the image. We found for the HRC data that focus was an important parameter to determine. For the WFPC2 data, the default focus value was adequate in most cases and no further adjustments were justified.

#### *3.2.3. Refined PSF fit*

We determined each of seven components, average background, position (x1, y1) and flux (f1) of the primary and position (x2, y2) and flux (f2) of the secondary individually by minimizing the $\chi^2$ of the residual between the model and the sub-image. Once all the position (x,y) and flux scaling (f) were refined we re-determined the average background. This process resulted in a refined initial guess that was sufficient to enable automated fitting to proceed in most cases. The automated fitting employed the *amoeba* (Press et al. 1992) routine, which performs multidimensional minimization of a function containing all our variables using the downhill simplex method, to optimize the fits. We ran the automated process alternating with the focusing routine until the $\chi^2$ converged. Typically, 4-5 iterations were required to reach a final PSF model.

#### *3.2.4. Photometry*

Before extracting photometry from the model images it was necessary to make an additional correction for the geometric distortion in the HRC data. The flat-field-corrected (*flt*) images were background subtracted and multiplied by a "pixel area map" (PAM) correction image prior to the PSF fitting. At the location on the HRC where our data fell the PAM correction is a factor of ~1.12. The PAM correction does not affect the fitted position and no similar correction was necessary for WFPC2 images.

We extracted photometry from the final PSF model for each component of the binary. Counts were summed within a 0.5 arcsecond aperture and converted to magnitude in the standard fashion, $M_{\text{half}} = -2.5\log_{10}(\text{counts/sec})$. This value was then corrected to an infinite aperture by adding an offset value, $\Delta M_{inf}$, from Table 5 in Sirianni et al. (2005). The photometric zero point, $M_{Zpt}$, was determined from the *photflam* and *photzpt* values in the image header. It is given in the standard Space Telescope magnitude system (STmag). We used *synphot* to convert from the STmag system to the standard Vega magnitude system (Vegamag, $M_{Vega}$). The conversions are $M_{Vega}$(F606W, WFPC2)= -0.2989, $M_{Vega}$(F814W, WFPC2)= -1.2376, $M_{Vega}$(F606W, HRC)= -0.2369, and $M_{Vega}$(F814W, WFPC2)= -1.28.

A correction for the location- and flux-dependent charge transfer efficiency (CTE) is also required for HST data. For the HRC we followed the process described in Riess (2003) and for WFPC2 the example of Dolphin (2000ab). The correction for CTE losses is on the order of -0.05 magnitudes for both instruments.

Uncertainties were estimated in two ways. For individual *flt* or *c0f* images we determined the changes in flux that caused a 1-sigma change in the $\chi^2$ residual. Additionally, for groups of exposures taken in a single 2-3 orbit visit we computed the standard deviation using the corrected two-pass algorithm (Press et al. 1992). We found that the standard deviation from multiple exposures was comparable to or a few hundreths magnitude larger than the uncertainty calculated from the individual exposures. In all cases we report the largest uncertainty obtained from these two methods. We note that short period lightcurves are one potential source of systematic "noise" in our photometry that might increase the standard deviation as discussed in Section 5.4.

### *3.2.5. Color correction*

In order to compare the HST measurements with other color values in the literature we converted from the native HST filter photometry to Johnson V and Cousins I magnitudes using *synphot*. The color corrections, $\Delta M_{STD}$ used a reddened Kurucz model solar spectrum convolved with the appropriate filter profiles. The reddening, E(B-V), was computed from the F606W–F814W color and ranged from 0.15–0.6. The color corrections were 0.20–0.40 magnitudes for F606W to Johnson V (V is fainter than F606W) and 0.01–0.05 magnitudes for F814W to Cousins I (I is fainter than F814W).

### *3.2.6. Photometry Summary*

Our final calibrated magnitude, $M_c$ in the measured filter (i.e. in the F606W filter $M_c = M_{F606W}$, in the F814W $M_c=M_{F814W}$) for each object listed in Table 2 is based on the terms found in Equation 1:

$$M_{\rm c} = M_{\rm half} + \Delta M_{\rm inf} + M_{\rm Zpt} + CTE + M_{\rm vega}. \quad (1)$$

Objects are sorted by the color of the primary component (within each native filter sample pair) and the second to last column tabulates the difference between the primary and secondary components. In most cases this difference is small, <0.1 magnitudes. The standard filter colors V and I, tabulated in Table 3 and again sorted relative to the color of the primary component, were calculated with the addition of the color correction factor, $F_{STD}$, appropriate for conversion from the respective native filters. In the Appendix we document a complete example of our photometry pipeline. We include the V-I colors for the combined light of the 6 unresolved binary systems in Table 3 since we find in Section 4 that the colors of the individual components are nearly identical.

**[ Table 2 ] and [ Table 3 ]**

Table 3 includes calculation of the spectral gradient, S, of our observations for more direct comparison with surveys in various filters. The spectral gradient is calculated following Boehnhardt et al. (2001). It is a measure of the reddening of the reflectance spectrum between two wavelengths, V and some other filter (in the case of this paper either R or I), expressed in percent of reddening per 100 nm. A value of S=0.0 means the object has exactly solar colors.

## 4. Statistical analysis

A plot of the color of the primary vs. the color of the secondary of each binary system is found in Figure 1. As can be seen by inspection, the colors of the components are clearly correlated. We quantified the correlation using two different statistical measures, the Spearman rank correlation and the Pearson product-moment coefficient. We describe each briefly.

**[ Figure 1 ]**

The Spearman rank correlation is a non-parametric statistical test that does not make any assumptions about the functional form of a possible correlation and does not assume a normal distribution of either variable. The independent variable, in this case the color of the primary, is ranked in order of increasing F606W-F814W color. The secondaries are similarly ranked and the correlation coefficient measures the difference between these two rankings. In a noiseless, perfectly correlated data set the rankings would be identical. Using the data in Table 2, we find a Spearman rank coefficient of $\rho = 0.8008$, corresponding to 99.99991% probability (5 sigma for a normal distribution) that the colors are correlated.

As a second check we computed the Pearson product-moment coefficient for the measured F606W – F814W colors. The Pearson product-moment coefficient applies to situations where the possible relationship between variables is linear. Other than linearity, there are no assumptions about the distribution, in particular it need not be normal. We find a Pearson product moment coefficient of $r = 0.84$. Using a probability table we find that the primary and secondary colors are correlated at a significance much more than 99%, in agreement with the Spearman rank test.

Two variables can be correlated, but systematically offset from one another. In order to determine whether there are any color offsets between primaries and secondaries in our sample we performed an additional test. In Figure 1 the dashed line is the locus of identical component colors which, by definition, goes through the origin. The best-fit line with a slope of 1 has a y-intercept of $0.01 \pm 0.01$ magnitudes. Thus, to the limit of our measurement uncertainties, the colors of primaries and secondaries are consistent with identical colors. We conclude that binary colors are not only correlated, but are statistically identical.

## 5. Discussion

### *5.1. Comparison to ground based photometry*

In Table 4 we list derived absolute magnitudes for the combined (primary+secondary) light of TNBs at zero phase angle, $H_V(1,1,0)$ or simply $H_V$. We also tabulate the spectral gradient, S (Boehnhardt et al. 2001), for these measurements. We compare both $H_V$ and S with published (or derived) values from other research teams. We include in this table the combined primary and secondary magnitudes for the four objects (listed in Section 2) that we did not resolve. We calculate $H_V$ following the methodology described by Bowell et al. (1989) and discussed with specific application to TNOs by Romanishin et al. (2005). Following these authors, we used G=0.15 for all the objects in the table. The details of the phase correction can be found at the end of Appendix A. We calculate S as described in Section 3.2.6. We use V-I and V-R color information from the sources referenced in the last column of Table 4.

Absolute magnitudes are published for all TNOs by the Minor Planet Center (MPC). We find, as have others (Romanishin & Tegler, 2005), that MPC magnitudes are in poor agreement with our photometry. In most cases we find that our values for $H_V$ are fainter than the MPC's with a median difference of 0.3 magnitudes. We suggest, similar to Romanishin & Tegler (2005), that disagreements with the MPC database are the result of $H_V$ magnitudes determined in the MPC from observations submitted from a variety of sites, transformed from a variety of filters and also based on measurements obtained for astrometric purposes instead of photometric accuracy.

**[ Table 4 ]**

Only 40% of our targets have reported photometric measurements in the literature. For those objects we find agreement with at least one literature measurement within the error bars for

~60% of our targets. However, we do find striking disagreement with some published values for a few of our targets; we comment on them briefly.

The largest disagreement we find is for (47171) 1999 $TC_{36}$. We found $H_V$=3.95±0.01, more than 1 magnitude brighter than three other published values. (47171) 1999 $TC_{36}$ is suspected to be a triple system (Jacobson 2007) and has, in the past, been found to have an inconsistent lightcurve (Peixinho et al. 2002; Ortiz et al. 2003). However, we have no definitive explanation for this large discrepancy.

We also find a significant difference for (80806) 2000 $CM_{105}$ where we measured $H_V$= 7.10±0.09, 0.8 magnitudes brighter than that found by Hainaut and Delsanti (2002). Lacerda et al. (2006) put an upper limit on variability in this object of 0.14 magnitudes from relative photometry taken over a period of several hours on each of two different nights. There is, thus, little to suggest that the large difference we observe might be due to lightcurve variations.

In a few other cases magnitude differences may be the result of observations obtained at different points in the lightcurves of the objects. (65489) Ceto/Phorcys has been reported to have a lightcurve with an amplitude of 0.13±0.02 magnitudes and a period of 4.43±0.03 hours (Barucci et al. 2008). (26308) 1998 $SM_{165}$ is known to have an 8.40±0.05 hour period with an amplitude >0.45 magnitudes (Spencer et al. 2006). (82075) 2000 $YW_{134}$, and (79360) 1997 $CS_{29}$ have been reported to have upper limits on variability of 0.10 magnitudes in observations taken sparsely over multiple nights (Sheppard 2003; Sheppard & Jewitt 2002). 1999 $OJ_4$ may have a lightcurve of a few tenths of a magnitude (Benecchi, unpublished data). Lastly, we note that (58534) Logos/Zoe gives a consistent $H_V$ with other published measurements, but is known to exhibit large variability (Noll et al. 2002).

With respect to ground-based spectral slope we find that our S values for the combined pairs are consistent with the S values determined for datasets in the literature within the error bars with a few exceptions. (58534) Logos, and (47171) $TC_{36}$ are both more neutral than literature values and (66652) Borasisi is more red. The later may be due, at least partially, to the color conversion offsets from F475W to V which are different than those used for the other objects we measured.

### *5.2. Comparison to single TNOs*

In Figure 2 we plot a comparison of the colors of TNB components with the colors of assumed single TNOs from the MBOSS (Hainaut & Delsanti 2002) and HST (Stephens 2007) color databases. The TNB components appear to span the same color range as single TNOs. In order to evaluate the significance of the apparent similarity of binary colors with singles we carried out a K-S test. Using all of the data shown in figure 2 we find a high probability of correlation, 96% (Figure 3). However there are possible biases in the data sample that could be important.

**[ Figure 2 ] and [ Figure 3 ]**

First, it is important to note that some fraction of the apparently single objects are likely to be binaries. For objects not yet observed by HST, it can be estimated that ~20% or more of Cold Classical objects and ~10% or fewer of the other dynamical classes (Noll et al. 2008a) are detectable binaries at HST WFPC2 resolution. However, a significant fraction of objects in MBOSS have already been searched for binaries with HST so the potential for resolvable binaries is limited. The distribution of binaries as a function of separation (Kern and Elliot, 2006; Noll et al. 2008a) suggests that a substantial fraction of binaries may exist, but be unresolvable by HST. Estimating the number of objects in the apparently single dataset is problematic. In the most extreme case, all TNOs could be binary. In that case, we would be comparing color as a function of binary separation.

A more tractable source of potential bias is related to the known correlation of color with dynamical class (Tegler and Romanishin 2000, Gulbis et al. 2006). In particular, the Cold Classical population is known to be systematically redder than other dynamical classes. Because binaries are more common in the Cold Classical population, our sample of binaries is skewed with 44% Cold Classicals compared to 27% for the singles. To try to account for this bias we randomly selected 10 samples of singles with the same mixture of Cold Classicals and non-Cold Classicals as in our binary data. We find that the correlation probability ranges from 60-96% in these ten tests. Any bias from the dynamical makeup of the sample appears to be small and if anything reduce the correlation. However, in general, the correlation remains high leading us to conclude that whatever mechanism is responsible for the colors of TNOs, it affects TNBs similarly.

### *5.3. Color Correlations*

How might the identical colors of the primary and the secondary in binaries have arisen? There are two possibilities. TNO colors could be the product of local environmental factors related to their current heliocentric orbit. In this case, because binary components share the same heliocentric orbits, the environmental effects related to their orbits are the same. Alternatively, identical colors may result if TNO colors are genetic. In this scenario, binary components have the same composition and color because each component accreted at the same location in the protoplanetary disk before becoming gravitationally bound. Finally, impacts can play a role both as an environmental factor and in the creation of bound systems from a single parent body. We discuss each of these different possibilities.

In addressing the origin of identical binary colors, it is worth keeping in mind the strong evidence that TNBs are primordial (*cf*. Noll et al. 2008a). System angular momentum indicates that dynamical capture is the only possible mode of formation for many TNBs. This formation mode requires a density much larger than currently exists in the Kuiper Belt (Goldreich et al. 2002). Similarly, collisional models that may account for the satellites of the largest TNOs, also require higher density of impactors than is found in the current Kuiper Belt (Canup 2005, Levison et al. 2007). Whether formed by capture or impact, binaries and multiples must have formed before the protoplanetary disk was depleted to its current anemic state.

*5.3.1. Current Environment*

With the exception of the giant-planet crossing TNBs (42355) Typhon/Echidna and (65489) Ceto/Phorcys, the known binaries are currently found in orbits that are stable on Gyr timescales. Thus, if environmental factors play a role in determining the colors of TNOs, it should be possible to identify relationships between dynamical status and color in ensembles of objects.

A number of investigations have searched for measurable trends of color with current dynamical status. The strongest color trend that has been identified to date is the trend with inclination among Classical TNOs (*cf*. Doressoundiram 2008; low inclination and eccentricity orbits). However, as shown by Thébault & Doressoundiram (2003), this trend cannot be explained by impact energy (Stern 2002). Instead, this trend has generally been understood as a result of Classicals being composed of two overlapping populations with intrinsically different colors. As argued by Noll et al. (2008a), the sharp difference in binaries between high and low inclination Classicals indicates that their physical differences, including color and albedo (Brucker et al. 2008; Grundy et al. 2005), are likely primordial.

The peak temperature of a TNO is a physical factor that might influence color. The peak temperature of a TNO is most closely related to its perihelion. However, equilibrium temperature varies only slowly, as $r^{-0.5}$, so that plausible temperature differences are on the order of 10K or less. Albedo variations can produce a temperature effect nearly as large, as can pole orientation and thermal inertia. All of these unknowns conspire to confound the interpretation of any observed trends. In existing color surveys, no strong trend of TNO color with perihelion has been observed (Doressoundiram et al. 2008).

Variation in cosmic ray flux and solar wind is another mechanism that might affect the surface properties of TNOs. Solar wind flux drops off as $1/r^2$ as expected. However, beyond ~10AU the particle density is dominated by neutrals that remain roughly constant with heliocentric distance (Richardson & Schwadron 2008). TNOs with aphelia greater than ~100AU may penetrate the asymmetric heliopause where they may be exposed to a flux of interstellar ions comparable in density to the neutrals. Cooper et al. (2003) calculate that the uppermost 10nm accumulates 100eV per atom in a relatively short ~$10^6$ yr timescale, although at depths of a micron or more the timescale is longer than the age of the solar system. Taken together, there is little to suggest that particle irradiation can account for differences in TNO colors.

Micrometeorite gardening is another environmental factor that plays a role in space weathering. The plasma wave detectors on Voyager 1 and 2 found impact rates out to about 30 and 50 AU, respectively, to be roughly constant (Grün et al. 2000), ruling out variations in micrometeorite impacts as a source of global variation in TNO surfaces. The stochastic nature of impact resurfacing (Jewitt & Luu 2001) makes it likely, however, that separate components in a binary would have different collisional histories and potentially strong color variation lightcurve signatures. The identity of binary component colors and lack of color variation (Section 5.4) is, therefore, evidence against impact resurfacing as a mechanism for creating the observed color diversity of TNOs.

For binaries we must also consider the unique microenvironment in which these objects reside. Stern (2008) has shown that for the largest TNOs, material ejected from the surfaces of small satellites remains gravitationally bound in the binary (or multiple) system and can lead to a shared surface regolith. However, as Stern also calculates, for the smaller TNOs in our sample, micrometeorite ejecta velocities exceed the escape velocity of the pair and the total mass accreted can be estimated simply as the total mass ejected by impacts times the fractional solid angle subtended by the object as seen from its companion.

Durda & Stern (2000) estimate the mass ejected by TNOs to be between 4–70×$10^{18}$ g $(r/100\ \mathrm{km})^2$, where r is the radius of the target body. Following Stern (2008), the fraction of this mass that can be intercepted ranges from $10^{-6}$ to $5\times10^{-4}$ for the systems with published binary orbits (Noll et al. 2008a, Table 2). This fraction results in mass per unit area ranging from 0.02 to 3.5 g cm$^{-2}$, or a layer 0.04 to 7 cm thick for density 0.5 g cm$^{-3}$. Stern points out that a layer this thick is enough to dominate the photometrically active surface. However, for this mechanism to work for the small, more equally sized binaries, one must postulate, implausibly, that the accumulating veneer of material from the companion body is undisturbed over the age of the solar system at the same time that $2{\times}10^3$ to $10^6$ times as much mass is being removed from the surface by impacts.

Observational tests of ejecta exchange are possible. Ejecta exchange in comparable mass binaries, like the ones that dominate our sample, would produce a color distribution with a smaller variance around the median of observed TNO colors than the general TNO population by "averaging" the component colors. However, as we describe in section 5.2, the color distribution of TNBs is a very close match to that of TNOs in general. We conclude that the TNB color distribution shows no evidence of ejecta exchange among the objects measured in our sample. Another possible test occurs in tidally locked binaries which would produce facing hemispheres with a common color while oppositely facing hemispheres would retain their original colors. At present, however, there are few known synchronous binaries and little data to constrain color-dependent lightcurves as we discuss in Section 5.4. Future observations may remedy this lack of data.

In summary, the comparability of TNB colors to the colors of apparently single TNOs argues that the influence of the current local environment, if any, on the colors of TNBs must be the same as it is for the general TNOs population. Thus, we conclude that there is little evidence that the environment in which TNBs have resided for the last 4 Gyr has had a determinative influence on their colors. The identity of binary component colors, is thus not explained by environmental factors.

#### *5.3.2. Composition*

If local environmental factors are not the source of color correlation in TNBs, then we must look to TNO genetics. Many TNOs, perhaps all, originated elsewhere in the protoplanetary disk than their current locations in the Kuiper Belt (Gomes et al. 2005; Morbidelli et al. 2005; Tsiganis et al. 2005; Levison et al. 2008). In some models, TNOs originated over a wide range of heliocentric distance from ~10 AU out to ~30 AU or more. Interestingly, this distance range spans several possible compositional thresholds, at increasing distances from the Sun, where hydrocarbons, methane, and carbon monoxide become stable solids (Ciesla et. al. 2006). TNO color diversity might then trace compositional differences in the protoplanetary disk, even allowing for subsequent space weathering.

Same-color binaries are a natural outcome if compositionally distinct zones were present in the protoplanetary disk. Because significantly higher densities are required for binary formation by both dynamical capture and collisions (Goldreich et al. 2002; Levison et al. 2007), binaries must have formed before any substantial mixing of the disk occurred. Thus, the components in each binary can be expected to have identical compositions determined by the physical properties of the protoplanetary disk where they coalesced. Color then, emerges as an observable proxy for primordial composition as has been argued elsewhere (Tegler et al. 2008; Noll et al. 2008b). Interestingly, the fact that TNB colors span the same range as TNO colors also allows us to conclude that binary formation occurred at all locations in the protoplanetary disk and not in a preferred locale.

Collisions may also play a minor role in creating bound systems with identical surface composition and colors. 2003 $EL_{61}$ and its associated family appear to be one such example (Brown et al. 2007; Levison et al. 2007). However, most of the objects in our sample cannot have originated in this way and so, we conclude, collisions cannot be responsible for the global color properties of TNBs.

### *5.4. Variability*

As part of our determination of the resolved colors of TNBs, we also assess the variability of the components. Two different approaches were followed.

First, we looked for evidence of short-period variability that might result from lightcurve variations. Because we calculated the uncertainties in our magnitudes based on the scatter of an object's measurements during the 90-180 minute HST observation, the presence of real variation in the brightness of one or both components of a TNB will result in an inflated photometric

uncertainty. At least one component of (148780) 2001 $UQ_{18}$, 2000 $QL_{251}$, 2001 $QY_{297}$ and 2001 $XR_{254}$ showed evidence of lightcurve variability during this timespan on the order of a few tenths of a magnitude. Further evaluation of the lightcurves of these objects will be addressed in a future paper after all the data for these objects have been obtained through our 11178 HST program. For now we note that the photometric uncertainty for these four objects may be partly inflated by lightcurve variations. The color measurements are consistent within the uncertainties across HST orbits indicating no substantial color variability.

Second, six TNBs had color measurements at multiple epochs (spanning as short as 24 days and as long as 419 days). We correct the magnitudes for different observational geometry (Appendix, Eq. A4) and plot in Figure 4 the color variation of the primary component vs. time scaled to the duration of the observations [(42355) Typhon $t_f$=306.380 days, (65489) Ceto $t_f$=24.757 days, 2003 $QY_{90}$ $t_f$=419.776 days, 2003 $TJ_{58}$ $t_f$=78.249 days, 1998 $WW_{31}$ $t_f$=226.333 days and (66652) Borasisi $t_f$=100.689 days]. In all cases the color of the secondary component is nearly identical to the primary so for clarity we only plot the primary component.

Overall, the colors of the TNOs are consistent between visits. In a few cases, the colors of the Centaurs and one of the TNOs, 1998 $WW_{31}$, there is variation off the median by slightly more (2-sigma) than the error bars (0.03 magnitudes) in one or more epochs, however with respect to the global statistics of the measurements, none of these variations are significant.

**[ Figure 4 ]**

## 6. Conclusions

We have measured the colors of both components of 23 TNBs. TNBs display a wide range of colors, but within individual binary systems colors are strongly consistent. We find that the colors of the components are correlated at the 99.99991% level based on a Spearman rank calculation. They are also identical to each other within the measurement uncertainties. Colors of TNBs cover the same range as apparently single TNOs suggesting that there are no special mechanisms affecting the colors of binaries that are not operative on singles. The distributions are indistinguishable at the 96% level with a K-S test. Four objects show evidence for short duration (90-180 minute) lightcurve variability, and three TNBs show hints of possible color variation, but the result is not strong.

We reviewed the possible environmental mechanisms that might produce correlated colors, but did not find any that could unambiguously produce correlated colors. The exchange of regolith caused by impacts (Stern 2008) does allow for the accretion of small amounts of material on companions, but is dwarfed by the loss of material and other resurfacing effects caused by impacts. In the absence of a convincing environmental mechanism, we conclude that the correlated colors of binaries can be best explained if colors of TNBs, and TNOs in general, are the surface expression of compositional differences set in the protoplanetary nebula. Color gradients in circumstellar disks may reveal analogous processes in other planet-forming environments.

## Appendix: HST Photometry

It became clear during our analysis that there was no coherent comprehensive reference for photometric analysis of our HST datasets. We relied on a number of internal HST documents, the instrument handbooks as well some papers in the literature. Therefore, we provide in this appendix a complete description of our photometric pipeline to document our analysis and provide a cohesive reference point.

The calibrated magnitude, $M_c$, is a combination of 6 factors for both instruments, one that is obtained by direct measurement from the image, while the other five are offsets to correct for various aspects of the instrumentation. The summed components include:

$$M_{\mathrm{c}} = M_{\mathrm{half}} + \Delta M_{\mathrm{inf}} + M_{\mathrm{Zpt}} + CTE + M_{\mathrm{vega}} + \Delta M_{\mathrm{STD}}. \tag{A1}$$

$M_{half}$ is the measured component, summed counts, from a circular aperture that extends out to 0.5 arcseconds placed around the source. Model PSFs created with TinyTim were shifted and scaled to match the flux of the primary and secondary in each binary system. The counts within a 0.5" aperture were then summed for the individual primary and secondary models that reproduced the data. Two main effects come from using the models instead of the raw data: (1) the object counts are not noisy since they come from a model, and (2) our measured signal is not complicated by the signal from the binary companion. $M_{inf}$ is read off of Table 5 in Sirianni et al. (2005) for the particular filter in which the observation is obtained for ACS/HRC and is 0.1 magnitudes for WFPC2 photometry irrespective of filter. It is a single factor that compensates for the additional flux of the object between 0.5" and nominal "infinity," the standard reference point for *synphot*. $M_{Zpt}$ is calculated from values provided in the image header keywords *photflam* and *photzpt* such that $\Delta M_{\mathrm{Zpt}} = -2.5\log(photflam) - photzpt$. However, this magnitude is still in the internal HST system, STmag, and needs to be converted to a standard stellar magnitude, referenced in *synphot* as the Vegamag system. We use a Kurucz solar model as our reference standard and find that the conversion factors, $M_{Vega}$, for F606W and F814W are –0.2989 and –1.2376, respectively for WFPC2. Likewise, the factors are -0.2369 and -1.28 for ACS/HRC.

Finally, there is a correction for where the object is located on the chip to account for deficits in charge transfer efficiency, CTE, an effect that causes some signal to be lost when charge is transferred down the chip during readout. The calculation for CTE is different for each instrument and the effect is more significant for WFPC2 than for ACS/HRC. To calculate CTE for WFPC2 we follow the example of Dolphin (2000ab) using the following equation set:

$$CTE = YCTE + XCTE = \frac{\ln\left(e^{c1} * [1 + c2] - c2\right)}{0.436} + 0.0021 * \frac{x}{800} e^{(-0.234 * bg)} \tag{A2}$$

where $c1 = 0.0114\left(0.67 * e^{(-0.246 * lbg)} + 0.330 * e^{(-0.0359 * bg)}\right) * \left(1 + 0.335 * yr - 0.0074 * yr^2\right) * \frac{y}{800}$,

$c2 = 3.55 e^{(-0.474 * lct)}$, $lct0 = \ln\left(\frac{counts}{electrons}\right) - 7$, $bg = \sqrt{\left[\frac{background}{electrons}\right]^2 + 1} - 10$,

$lbg = \ln\left(\sqrt{\left[\frac{background}{electrons}\right]^2 + 1}\right) - 1$, $yr = \frac{(MJD - 50193)}{365.25}$, and $lct = lct0 + 0.921 * XCTE$.

For ACS/HRC we follow the process described by Riess (2003) and the equation set:

$$CTE = 10^{-0.89} \times SKY^{-0.24} \times FLUX^{-0.21} \times \frac{Y_{pos}}{1024} \times \frac{(MJD - 52333)}{(52714 - 52333)} \tag{A3}$$

with the exponents for the first three parameters based on a 3 pixel aperture. The result of all of these components is the calibrated magnitude of the object in the observed HST filter in the standard Vegamag system.

For comparison with the standard Johnson-Cousins colors in the literature we perform one more conversion, from F606W to Johnson V and F814W to Cousins I. The values are different for each camera. The conversion actually requires two calculations since we want to find the offset using a solar spectrum of similar reddening as the object itself. Using *synphot* we create a number of look-up tables. We construct a table of F606W-F814W values vs. the reddening function (ebmv) for our Kurucz model (call this table 1) and a table of embv vs. filter conversions (F606-Johnson V, F814-Cousins I etc.; call this table 2). For a given object we take our measured F606W-F814W then use the table to look up the appropriate ebmv value for that object. We then find the offset values between the observed and desired filters with the Kurucz model reddened appropriately for the object. We add the offset value to the predetermined calibrated magnitude for our object in the HST filter system to find the actual magnitude of the object in standard filters. For example, F606W = 22.26 and F814W = 21.43 so F606W-F814W = 0.83. From table 1 we find embv = 0.4. From table 2 we find that $\Delta M_{STD}$ = 0.345 for F606W to Johnson V and $\Delta M_{STD}$ = 0.007 for F814W to Cousins I at this ebmv value. Our final magnitudes are: V = 22.60 and I = 21.43 so V-I=1.17. To determine the uncertainties we consider the minimum and maximum deltas in the natural HST filters (found from table 1) to identify the range of magnitude offsets in table 2 and propagate the uncertainties from there. The uncertainty in V grows more than the uncertainty in I since the shift in wavelength is larger from F606W to Johnson V than F814W to Cousins I.

Finally, so that our measurements can easily be compared with observations under different geometric circumstances, we calculate the H-magnitude (the mean magnitude of a small body in the absence of rotational or aspect variations at a solar phase angle α, reduced to heliocentric and geocentric distances of 1AU) in the respective filters and record them in Table 2. We follow the methodology of Bowell et al. (1989) as applied to the Kuiper Belt following the pattern of Romanishin & Tegler (2005). In Equation A4 *r* is heliocentric distance (AU), Δ is geocentric distance (AU), and α is the sun-target-observer angle (degrees) at the time of observation. We don't have information for constraining *G* so we use the value adopted in the literature of 0.15 (Romanishin & Tegler 2005),

$$H_{filter} = M_{filter} - 5\log(r\Delta) + 2.5\log\left[(1-G)\Phi_1(\alpha) + G\Phi_2(\alpha)\right] \text{ where}$$
$$\phi_1(\alpha) = \exp\left[-3.33\left(\tan\frac{1}{2}\alpha\right)^{0.63}\right] \text{ and } \phi_2(\alpha) = \exp\left[-1.87\left(\tan\frac{1}{2}\alpha\right)^{1.22}\right]. \tag{A4}$$

**Acknowledgments**

This work is based on observations made with the NASA/ESA Hubble Space Telescope. These observations are associated with programs # 9991, and 11178. Support for these programs was provided by NASA through a grant from the Space Telescope Science Institute, which is operated by the Association of Universities for Research in Astronomy, Inc., under NASA contract NAS 5-26555. We also wish to thank two anonymous reviewers and A. Stern for a prepublication copy of his 2008 paper.

## Tables

TABLE 1. LOG OF OBSERVATIONS

| PropID | PI | No. obj | Camera | Filters/Observing Sequence |
|---|---|---|---|---|
| 11178 | W. M. Grundy | 12 | WFPC2 | 2xF606W (260s), 4xF814W (500s), 2xF606W (260s) dithered |
| 10508 | W. M. Grundy | 3 | ACS/HRC | 2xF606W (223s), 4xF814W (223s), 2xF606W (223s) dithered |
| 9991 | K. S. Noll | 1 | ACS/HRC | 2xF775W (500s), 2xF475W (600s) |
| 9746 | J. L. Margot | 4 | ACS/HRC | 2xF606W (610s), 2xF814W (610s) |
| 9508 | C. Veillet | 1 | WFPC2 | Various combinations of F555W (500, 600s), F675W (500, 700s) and F814W (500, 600s) |
| 9259 | C. Veillet | 1 | WFPC2 | 2xF555W (230s), 2xF675W (230s), 2xF814W (300s) |
| 9320 | C. Veillet | 1 | WFPC2 | 2xF555W (230s), 2xF675W (230s), 2xF814W (300s) |

TABLE 2. RESOLVED HST MAGNITUDES

| Object | MJD‡ | Separation (arcsec) | ΔM F606W | Primary F606W | Primary F814W | Primary Color§ | Secondary F606W | Secondary F814W | Secondary Color§ | Δ Color§ | Ref |
|---|---|---|---|---|---|---|---|---|---|---|---|
| 2000 $QL_{251}$ | 54339.75743 | 0.267±0.001 | 0.27±0.09 | 23.36±0.07 | 22.76±0.03 | 0.60±0.07 | 23.63±0.06 | 22.90±0.09 | 0.73±0.10 | -0.13±0.13 | 1 |
| (120347) 2004 $SB_{60}$† | 54324.32378 | 0.156±0.002 | 2.19±0.02 | 20.66±0.01 | 20.04±0.01 | 0.62±0.01 | 22.85±0.02 | 22.22±0.04 | 0.63±0.04 | -0.01±0.04 | 1 |
| 2001 $QC_{298}$ | 53175.67839 | 0.108±0.001 | 0.51±0.04 | 23.07±0.02 | 22.45±0.03 | 0.62±0.04 | 23.58±0.03 | 22.90±0.03 | 0.68±0.04 | -0.07±0.06 | 3 |
| (60458) 2000 $CM_{114}$† | 54483.41189 | 0.082±0.002 | 0.60±0.08 | 23.90±0.04 | 23.29±0.08 | 0.62±0.09 | 24.50±0.07 | 23.75±0.08 | 0.75±0.11 | -0.13±0.14 | 1 |
| 2001XR $_{254}$ | 54439.41264 | 0.105±0.001 | 0.07±0.09 | 23.06±0.05 | 22.32±0.07 | 0.74±0.09 | 23.12±0.07 | 22.36±0.07 | 0.76±0.10 | -0.02±0.13 | 1 |
| (42355) Typhon | 53780.55666 | 0.169±0.001 | 1.32±0.03 | 20.05±0.03 | 19.23±0.01 | 0.83±0.03 | 21.38±0.02 | 20.61±0.02 | 0.77±0.02 | 0.06±0.04 | 2 |
| | 53785.07234 | 0.160±0.001 | 1.36±0.04 | 20.11±0.02 | 19.35±0.05 | 0.77±0.06 | 21.47±0.03 | 20.78±0.02 | 0.70±0.04 | 0.07±0.07 | 2 |
| | 54044.06850 | 0.135±0.001 | 1.05±0.03 | 20.62±0.03 | 19.91±0.01 | 0.72±0.03 | 21.68±0.02 | 20.98±0.01 | 0.69±0.02 | 0.02±0.03 | 2 |
| | 54091.45205 | 0.102±0.001 | 1.23±0.04 | 20.55±0.02 | 19.77±0.01 | 0.78±0.02 | 21.78±0.03 | 21.01±0.02 | 0.77±0.04 | 0.01±0.04 | 2 |
| (134860) 2000 $OJ_{67}$ | 54319.38434 | 0.073±0.001 | 0.82±0.05 | 22.88±0.03 | 22.08±0.05 | 0.80±0.06 | 23.70±0.04 | 22.66±0.03 | 1.04±0.05 | -0.24±0.07 | 1 |
| $2004PB_{108}$ | 54375.28295 | 0.300±0.003 | 1.11±0.09 | 23.56±0.04 | 22.76±0.05 | 0.80±0.06 | 24.67±0.08 | 23.83±0.09 | 0.84±0.12 | -0.04±0.13 | 1 |
| (58534) Logos | 53179.04682 | 0.072±0.001 | 0.33±0.06 | 23.48±0.04 | 22.65±0.05 | 0.83±0.07 | 23.81±0.04 | 22.93±0.17 | 0.88±0.17 | -0.04±0.18 | 3 |
| (65489) Ceto | 53861.31526 | 0.084±0.002 | 0.57±0.03 | 21.23±0.02 | 20.30±0.09 | 0.94±0.09 | 21.80±0.03 | 20.97±0.02 | 0.83±0.03 | 0.11±0.10 | 2 |
| | 53865.18174 | 0.087±0.001 | 0.60±0.03 | 21.14±0.03 | 20.35±0.02 | 0.79±0.03 | 21.75±0.01 | 20.92±0.03 | 0.82±0.03 | -0.03±0.04 | 2 |
| | 53885.00662 | 0.083±0.001 | 0.54±0.09 | 21.31±0.02 | 20.44±0.01 | 0.87±0.02 | 21.85±0.08 | 21.09±0.02 | 0.76±0.09 | 0.11±0.09 | 2 |
| | 53886.07239 | 0.087±0.001 | 0.60±0.02 | 21.28±0.01 | 20.47±0.01 | 0.82±0.02 | 21.88±0.02 | 21.10±0.02 | 0.78±0.03 | 0.04±0.03 | 2 |
| 2003 $TJ_{58}$ | 54324.06038 | 0.090±0.002 | 0.44±0.09 | 24.02±0.05 | 23.06±0.06 | 0.96±0.08 | 24.46±0.08 | 23.74±0.07 | 0.72±0.10 | 0.24±0.13 | 1 |
| | 54346.09499 | 0.133±0.001 | 0.51±0.09 | 24.01±0.06 | 23.15±0.04 | 0.86±0.07 | 24.51±0.08 | 23.60±0.06 | 0.92±0.10 | -0.06±0.12 | 1 |
| | 54395.70344 | 0.171±0.001 | 0.52±0.10 | 23.88±0.05 | 23.01±0.04 | 0.87±0.07 | 24.41±0.08 | 23.50±0.06 | 0.90±0.10 | -0.03±0.12 | 1 |
| | 54424.24945 | 0.134±0.001 | 0.55±0.08 | 23.81±0.04 | 22.90±0.03 | 0.90±0.05 | 24.36±0.07 | 23.43±0.06 | 0.93±0.09 | -0.03±0.10 | 1 |
| 2003 $QW_{111}$ | 54682.80058 | 0.169±0.002 | 1.38±0.17 | 23.55±0.04 | 22.69±0.03 | 0.86±0.05 | 24.92±0.14 | 23.96±0.09 | 0.97±0.16 | -0.11±0.16 | 1 |
| | 54698.64919 | 0.284±0.001 | 1.42±0.15 | 23.57±0.04 | 22.60±0.04 | 0.97±0.07 | 24.99±0.15 | 24.09±0.09 | 0.90±0.17 | 0.07±0.17 | 1 |
| 1999 $OJ_4$ | 54336.36316 | 0.107±0.001 | 0.15±0.22 | 23.56±0.06 | 22.67±0.07 | 0.89±0.09 | 23.70±0.21 | 22.69±0.03 | 1.01±0.21 | -0.12±0.23 | 1 |
| (79360) $1997CS_{29}$ | 54457.58469 | 0.089±0.001 | 0.19±0.03 | 22.37±0.03 | 21.47±0.01 | 0.90±0.03 | 22.57±0.02 | 21.60±0.04 | 0.96±0.04 | -0.07±0.05 | 1 |
| (123509) 200 $WK_{183}$† | 54389.84163 | 0.068±0.003 | 0.05±0.05 | 23.58±0.03 | 22.66±0.02 | 0.92±0.04 | 23.62±0.03 | 22.76±0.03 | 0.86±0.05 | 0.06±0.06 | 1 |
| (148780) 2001 $UQ_{18}$ | 54416.37948 | 0.157±0.001 | 0.45±0.13 | 23.19±0.10 | 22.27±0.08 | 0.92±0.13 | 23.64±0.08 | 22.63±0.11 | 1.01±0.14 | -0.09±0.18 | 1 |
| (47171) 1999 $TC_{36}$ | 53594.96925 | 0.282±0.001 | 2.21±0.02 | 20.06±0.01 | 19.11±0.01 | 0.95±0.01 | 22.27±0.01 | 21.40±0.02 | 0.88±0.02 | 0.07±0.03 | 3 |
| 2000 $CF_{105}$ | 53155.17593 | 0.626±0.001 | 0.70±0.16 | 24.02±0.11 | 23.05±0.05 | 0.97±0.12 | 24.72±0.12 | 23.81±0.05 | 0.92±0.13 | 0.05±0.18 | 3 |
| 2000 $CQ_{114}$ | 54548.37046 | 0.196±0.001 | 0.12±0.19 | 24.16±0.04 | 23.16±0.05 | 0.99±0.06 | 24.27±0.18 | 23.32±0.06 | 1.04±0.15 | 0.05±0.22 | 1 |

| Object | MJD‡ | Separation (arcsec) | ΔM F606W | Primary F606W | Primary F814W | Primary Color§ | Secondary F606W | Secondary F814W | Secondary Color§ | Δ Color§ | Ref |
|---|---|---|---|---|---|---|---|---|---|---|---|
| 2001 $QY_{297}$ | 54586.87578 | 0.173±0.016 | 0.21±0.27 | 22.91±0.15 | 21.85±0.10 | 1.05±0.18 | 23.11±0.22 | 22.30±0.14 | 0.82±0.26 | 0.23±0.32 | 1 |
| 2003 $QY_{90}$ | 53576.09906 | 0.363±0.002 | 0.02±0.13 | 23.87±0.04 | 22.78±0.16 | 1.10±0.16 | 23.89±0.12 | 22.85±0.09 | 1.04±0.15 | 0.05±0.22 | 2 |
| | 53596.96684 | 0.358±0.001 | -0.11±0.06 | 23.88±0.05 | 22.82±0.05 | 1.06±0.07 | 23.77±0.03 | 22.69±0.05 | 1.08±0.06 | -0.01±0.09 | 2 |
| | 53617.48881 | 0.337±0.002 | 0.41±0.11 | 23.87±0.02 | 22.84±0.04 | 1.04±0.04 | 24.28±0.11 | 23.14±0.09 | 1.14±0.14 | -0.11±0.15 | 2 |
| | 53995.87532 | 0.195±0.003 | 0.17±0.12 | 23.86±0.02 | 22.76±0.05 | 1.09±0.05 | 24.03±0.12 | 23.07±0.04 | 0.95±0.12 | 0.14±0.13 | 2 |
| (26308) 1998 $SM_{165}$ | 53921.93540 | 0.295±0.001 | 2.20±0.05 | 21.63±0.02 | 20.51±0.02 | 1.12±0.03 | 23.83±0.04 | 22.79±0.01 | 1.04±0.04 | 0.08±0.05 | 3 |
| **Object** | **MJD‡** | **Separation (arcsec)** | **ΔM F555W** | **Primary F555W** | **Primary F814W** | **Primary Color§** | **Secondary F555W** | **Secondary F814W** | **Secondary Color§** | **Δ Color§** | **Ref** |
| 1998 $WW_{31}$ | 52102.88866 | 0.881±0.002 | 0.49±0.19 | 24.04±0.13 | 22.91±0.06 | 1.13±0.14 | 24.53±0.14 | 23.59±0.08 | 0.94±0.16 | 0.19±0.21 | 5 |
| | 52130.78102 | 0.783±0.008 | 0.17±0.02 | 24.17±0.01 | 23.33±0.24 | 0.84±0.24 | 24.34±0.01 | 23.04±0.38 | 1.31±0.38 | -0.47±0.45 | 5 |
| | 52162.67546 | 0.589±0.002 | -0.07±0.05 | 24.45±0.03 | 23.09±0.09 | 1.37±0.09 | 24.38±0.04 | 23.61±0.16 | 0.77±0.16 | 0.60±0.19 | 5 |
| | 52273.47616 | 0.416±0.002 | 0.64±0.05 | 23.82±0.03 | 22.97±0.03 | 0.84±0.04 | 24.45±0.05 | 23.34±0.15 | 1.11±0.16 | -0.27±0.16 | 6 |
| | 52293.51297 | 0.562±0.002 | 0.40±0.02 | 24.03±0.01 | 23.04±0.03 | 0.99±0.03 | 24.42±0.02 | 23.69±0.03 | 0.74±0.04 | 0.25±0.05 | 6 |
| | 52329.22130 | 0.767±0.003 | 0.76±0.24 | 24.03±0.08 | 22.93±0.17 | 1.10±0.18 | 24.78±0.23 | 23.61±0.04 | 1.17±0.23 | -0.08±0.30 | 6 |
| **Object** | **MJD‡** | **Separation (arcsec)** | **ΔM F475W** | **Primary F475W** | **Primary F775W** | **Primary Color§** | **Secondary F475W** | **Secondary F775W** | **Secondary Color§** | **Δ Color§** | **Ref** |
| (66652) Borasisi | 52871.61183 | 0.075±0.001 | 0.55±0.30 | 23.29±0.09 | 21.50±0.05 | 1.79±0.10 | 23.84±0.29 | 21.81±0.03 | 2.03±0.29 | -0.24±0.31 | 4 |
| | 52897.24733 | 0.228±0.001 | 0.52±0.08 | 23.34±0.05 | 21.49±0.03 | 1.85±0.06 | 23.87±0.07 | 21.88±0.06 | 1.99±0.09 | -0.14±0.11 | 4 |
| | 52960.50017 | 0.103±0.001 | 0.47±0.09 | 23.44±0.03 | 21.68±0.01 | 1.76±0.04 | 23.91±0.08 | 22.06±0.02 | 1.85±0.08 | -0.09±0.09 | 4 |
| | 52972.30088 | 0.079±0.001 | 0.45±0.04 | 23.38±0.02 | 21.65±0.05 | 1.72±0.05 | 23.82±0.03 | 22.05±0.01 | 1.77±0.03 | -0.05±0.06 | 4 |
| **Object** | **MJD‡** | **Separation (arcsecc)** | **ΔM Sloan *r’*** | **Primary Sloan *r’*** | **Primary Sloan *i*'** | **Primary Color§** | **Secondary Sloan *r’*** | **Secondary Sloan *i*'** | **Secondary Color§** | **Δ Color§** | **Ref** |
| (88611) Teharon∞ | 53261.13940 | 0.712±0.005 | 0.56±0.02 | 22.79±0.01 | 22.34±0.02 | 0.45±0.02 | 23.35±0.02 | 22.91±0.03 | 0.44±0.03 | 0.01±0.03 | 7 |
| 2005 $EO_{304}$ | 53849.68608 | 2.069±0.009 | 1.53±0.12 | 22.69±0.08 | 21.81±0.10 | 0.88±0.10 | 24.22±0.12 | 23.40±0.13 | 0.82±0.13 | 0.06±0.13 | 7 |

(1) Program 11178, WFPC2; (2) Program 10508, ACS/HRC; (3) Program 9746, ACS/HRC;
(4) Program 9991, ACS/HRC; (5) Program 9259, WFPC2; (6) Program 9230, WFPC2; (7) Ground based observations from Magellan with MagIC; ‡ MJD = JD-2400000.5, not light time corrected. § Color is defined with respect to the two filters listed, i.e. F606W-F814W, F555W-F184W, F475W-F775W, and Sloan *r’*-Sloan *i*'. † secondary/primary ambiguous. ∞ shorthand for Teharonhiawako.

TABLE 3. TRANSFORMED JOHNSON/COUSINS MAGNITUDES AND COLORS

| Object | Primary | | | | Secondary | | | | Δ (V-I) | Type† |
|---|---|---|---|---|---|---|---|---|---|---|
| | $H_V$‡ | $H_I$‡ | (V-I) | S§ | $H_V$‡ | $H_I$‡ | (V-I) | S§ | | |
| 2001 $QC_{298}$ | 8.18±0.03 | 7.40±0.04 | 0.78±0.05 | 1±2 | 8.69±0.04 | 7.84±0.03 | 0.85±0.05 | 1±2 | -0.07±0.07 | SD |
| Teharon | 6.48±0.02 | 5.66±0.03 | 0.82±0.04 | 1±2 | 7.04±0.03 | 6.23±0.04 | 0.81±0.05 | 1±2 | 0.01±0.06 | CL |
| 2000 $QL_{251}$ | 7.63±0.08 | 6.78±0.03 | 0.85±0.09 | 2±5 | 7.90±0.07 | 6.92±0.09 | 0.98±0.12 | 7±5 | -0.13±0.15 | 2:1 |
| (120347) | 4.39±0.01 | 3.52±0.01 | 0.87±0.01 | 2±1 | 6.58±0.02 | 5.70±0.04 | 0.89±0.04 | 3±2 | -0.02±0.04 | SD |
| (60458) | 7.91±0.06 | 7.03±0.08 | 0.87±0.10 | 2±5 | 8.50±0.10 | 7.50±0.08 | 1.00±0.13 | 8±6 | -0.13±0.16 | SD |
| 1998 $WW_{31}$ | 7.18±0.01 | 6.21±0.02 | 0.91±0.02 | 6±1 | 7.48±0.01 | 6.78±0.02 | 0.76±0.03 | 1±1 | 0.15±0.04 | CL |
| Typhon | 8.02±0.01 | 7.06±0.01 | 0.96±0.02 | 7±1 | 9.21±0.01 | 8.27±0.01 | 0.93±0.02 | 5±1 | 0.03±0.03 | CN |
| 2001 $XR_{254}$ | 6.82±0.08 | 5.78±0.07 | 1.04±0.11 | 10±5 | 6.89±0.10 | 5.82±0.07 | 1.06±0.12 | 11±6 | -0.02±0.16 | CL |
| Ceto | 6.94±0.01 | 5.89±0.01 | 1.05±0.02 | 10±1 | 7.50±0.01 | 6.51±0.01 | 1.01±0.02 | 8±1 | 0.04±0.03 | CN |
| Logos | 7.31±0.06 | 6.26±0.06 | 1.05±0.08 | 10±4 | 7.64±0.05 | 6.55±0.17 | 1.09±0.18 | 12±8 | -0.04±0.20 | CL |
| (134860) | 6.87±0.04 | 5.75±0.05 | 1.12±0.07 | 14±3 | 7.69±0.06 | 6.33±0.03 | 1.36±0.06 | 28±3 | -0.24±0.09 | CL |
| 2004 $PB_{108}$ | 7.45±0.05 | 6.33±0.05 | 1.12±0.07 | 14±4 | 8.56±0.10 | 7.40±0.09 | 1.16±0.13 | 16±6 | -0.04±0.15 | SD |
| 2003 $TJ_{58}$ | 7.92±0.03 | 6.79±0.02 | 1.13±0.04 | 14±2 | 8.43±0.04 | 7.32±0.03 | 1.11±0.05 | 13±3 | 0.02±0.06 | CL |
| (47171) | 4.08±0.01 | 2.89±0.01 | 1.19±0.01 | 18±1 | 6.30±0.02 | 5.18±0.02 | 1.12±0.03 | 14±1 | 0.07±0.03 | 3:2 |
| 2001 $FL_{185}$ | 7.41±0.21∞ | 6.22±0.10∞ | 1.19±0.23∞ | 18±12 | — | — | — | — | — | CL |
| 2000 $CF_{105}$ | 7.84±0.14 | 6.63±0.06 | 1.21±0.15 | 19±7 | 8.55±0.15 | 7.38±0.06 | 1.16±0.16 | 16±8 | 0.05±0.22 | CL |
| 1999 $OJ_4$ | 8.15±0.09 | 6.90±0.08 | 1.25±0.11 | 21±6 | 8.29±0.24 | 6.92±0.03 | 1.37±0.24 | 28±11 | -0.12±0.26 | CL |
| (82075) | 4.85±0.10∞ | 3.60±0.15∞ | 1.25±0.18∞ | 21±9 | — | — | — | — | — | 8:3 |
| (79360) | 6.27±0.04 | 5.00±0.01 | 1.26±0.04 | 22±2 | 6.46±0.03 | 5.13±0.04 | 1.33±0.05 | 26±2 | -0.07±0.06 | CL |
| 2003 $QW_{111}$ | 7.40±0.05 | 8.81±0.09 | 1.27±0.04 | 23±2 | 6.11±0.02 | 7.50±0.06 | 1.31±0.12 | 25±5 | -0.04±0.12 | 7:4 |
| (119979) | 4.88±0.07∞ | 3.60±0.06∞ | 1.28±0.09∞ | 23±4 | — | — | — | — | — | SD |
| (123509) | 7.51±0.05 | 6.22±0.02 | 1.30±0.05 | 24±3 | 7.56±0.04 | 6.32±0.03 | 1.24±0.06 | 20±3 | 0.06±0.08 | CL |
| (148780) | 6.98±0.14 | 5.69±0.08 | 1.30±0.16 | 24±8 | 7.43±0.12 | 6.05±0.11 | 1.39±0.17 | 29±8 | -0.09±0.24 | CL |
| 2003 $QY_{90}$ | 7.61±0.02 | 6.29±0.03 | 1.34±0.04 | 25±2 | 7.65±0.04 | 6.41±0.03 | 1.34±0.06 | 22±3 | 0.00±0.07 | CL |
| (80806) | 7.10±0.09∞ | 5.75±0.05∞ | 1.34±0.10∞ | 26±5 | — | — | — | — | — | CL |
| (182933) | 6.99±0.06∞ | 5.64±0.02∞ | 1.35±0.06∞ | 27±2 | — | — | — | — | — | SD |
| 1999 $RT_{214}$ | 8.13±0.02∞ | 6.73±0.06∞ | 1.39±0.06∞ | 30±2 | — | — | — | — | — | CL |
| (26308) | 6.11±0.03 | 4.71±0.02 | 1.40±0.04 | 30±2 | 8.31±0.05 | 6.99±0.01 | 1.32±0.05 | 25±2 | 0.08±0.06 | 2:1 |
| 2000 $CQ_{114}$ | 7.92±0.05 | 6.52±0.05 | 1.40±0.08 | 30±4 | 8.04±0.20 | 6.68±0.07 | 1.36±0.21 | 27±10 | 0.04±0.22 | CL |
| 2005 $EO_{304}$ | 6.57±0.09 | 5.14±0.11 | 1.43±0.14 | 32±6 | 8.09±0.13 | 6.73±0.14 | 1.36±0.19 | 27±8 | 0.07±0.24 | CL |
| Borasisi | 6.69±0.01 | 5.31±0.01 | 1.43±0.03 | 28±1 | 7.15±0.02 | 5.68±0.01 | 1.47±0.03 | 35±1 | -0.04±0.04 | CL |
| 2001 $QY_{297}$ | 6.82±0.22 | 5.34±0.10 | 1.48±0.24 | 36±12 | 7.03±0.28 | 5.78±0.15 | 1.25±0.31 | 21±16 | 0.23±0.39 | CL |

‡ Magnitude geometrically corrected (Appendix, Equation A4) to account for heliocentric, geocentric distance, phase angle of observation and phase coefficient; § Spectral gradient, a measure of the reddening of the reflectance spectrum between two wavelengths, V and I, expressed in percent of reddening per 100 nm; † CL=Classical, CN=Centaur, SD=Scattered. Derived from the Deep Ecliptic Survey (DES) database, Elliot et al. (2005). ∞ combined light values, the components were not uniquely resolved.

TABLE 4. GROUNDBASED AND HST COMPARISON

| Object | HST | | MPC | Literature | | |
|---|---|---|---|---|---|---|
| | $H_V$‡ | S§ | $H_V$ | $H_V$‡ | S§ | Ref |
| 2001 $QC_{298}$ | 7.66±0.03 | 1±2 | 6.1 | — | — | — |
| Teharon | 5.97±0.03 | 1±2 | 5.5 | — | — | — |
| 2000 $QL_{251}$ | 7.02±0.07 | 5±5 | 6.3 | — | — | — |
| (120347) | 4.26±0.02 | 2±1 | 4.4 | — | — | — |
| (60458) | 7.43±0.10 | 6±7 | 6.8 | 6.97±0.04 | 13±2 | 2 |
| | | | | 7.36±0.02 | 1±1 | 3 |
| 1998 $WW_{31}$ | 6.60±0.01 | 6±1 | 6.1 | — | — | — |
| Typhon | 7.75±0.01 | 9±1 | 7.2 | 7.68±0.04 | 12±2 | 1 |
| | | | | 7.65±0.02 | 0±0 | 3 |
| 2001 $XR_{254}$ | 6.10±0.04 | 10±3 | 5.6 | — | — | — |
| Ceto | 6.43±0.01 | 9±1 | 6.3 | 6.60±0.03 | 4±3 | 3 |
| Logos | 6.72±0.06 | 11±6 | 6.6 | 6.76±0.18 | 34±7 | 2 |
| (134860) | 6.49±0.06 | 20±3 | 6.0 | — | — | — |
| 2004 $PB_{108}$ | 7.13±0.12 | 15±8 | 6.3 | — | — | — |
| 2003 $TJ_{58}$ | 7.39±0.06 | 14±3 | 7.8 | — | — | — |
| (47171) | 3.95±0.01 | 17±1 | 4.9 | 5.272±0.055 | 47±3 | 1 |
| | | | | 4.86±0.06 | 32±2 | 2 |
| | | | | 5.33±0.02 | 16±1 | 3 |
| 2001 $FL_{185}$ | 7.41±0.21 | 18±12 | 7.0 | — | — | — |
| 2000 $CF_{105}$ | 7.38±0.14 | 18±8 | 6.9 | 7.59 | — | 3 |
| 1999 $OJ_4$ | 7.49±0.27 | 26±14 | 7.0 | 6.91±0.06 | 27±4 | 2 |
| (82075) | 4.85±0.10 | 21±9 | 5.0 | 4.39±0.06 | 19±4 | 2 |
| | | | | 4.74±0.02 | 3±3 | 3 |
| (79360) | 5.62±0.02 | 24±1 | 5.1 | 5.06±0.09 | 28±3 | 2 |
| | | | | 5.52±0.07 | 8±4 | 3 |
| 2003 $QW_{111}$ | 7.14±0.10 | 24±5 | 6.4 | — | — | — |
| (119979) | 4.88±0.07 | 23±4 | 5.1 | — | — | — |
| (123509) | 6.77±0.06 | 21±4 | 6.5 | — | — | — |
| (148780) | 6.44±0.07 | 26±5 | 5.6 | — | — | — |
| 2003 $QY_{90}$ | 6.94±0.03 | 26±2 | 6.3 | — | — | — |
| (80806) | 7.10±0.09 | 26±5 | 6.3 | 6.30±0.03 | 51±30 | 2 |
| (182933) | 6.99±0.06 | 27±2 | 6.4 | — | — | — |
| 1999 $RT_{214}$ | 8.13±0.02 | 30±2 | 7.8 | — | — | — |
| (26308) | 5.98±0.02 | 30±1 | 5.8 | 5.80±0.19 | 32±5 | 2 |
| | | | | 6.13±0.10 | 23±7 | 3 |
| 2000 $CQ_{114}$ | 7.21±0.25 | 28±13 | 6.8 | — | — | — |
| 2005 $EO_{304}$ | 6.33±0.15 | 17±10 | 6.3 | — | — | — |
| Borasisi | 6.25±0.01 | 39±1 | 5.9 | — | 22±3 | 4 |
| 2001 $QY_{297}$ | 6.09±0.03 | 25±2 | 5.4 | — | 20±3 | 4 |

(1) Rabinowitz et al. 2007; (2) Hainaut & Delsanti 2002; (3) Tegler et al. 2007; (4) Gulbis et al. 2006. ‡ Magnitude of the combined flux of the binary system (primary+secondary) geometrically corrected (Appendix, Equation A4) to account for heliocentric, geocentric distance, phase angle of observation and phase coefficient. § Spectral gradient, a measure of the slope of the reflectance spectrum expressed in percent per 100 nm (reference).

**FIGURES**

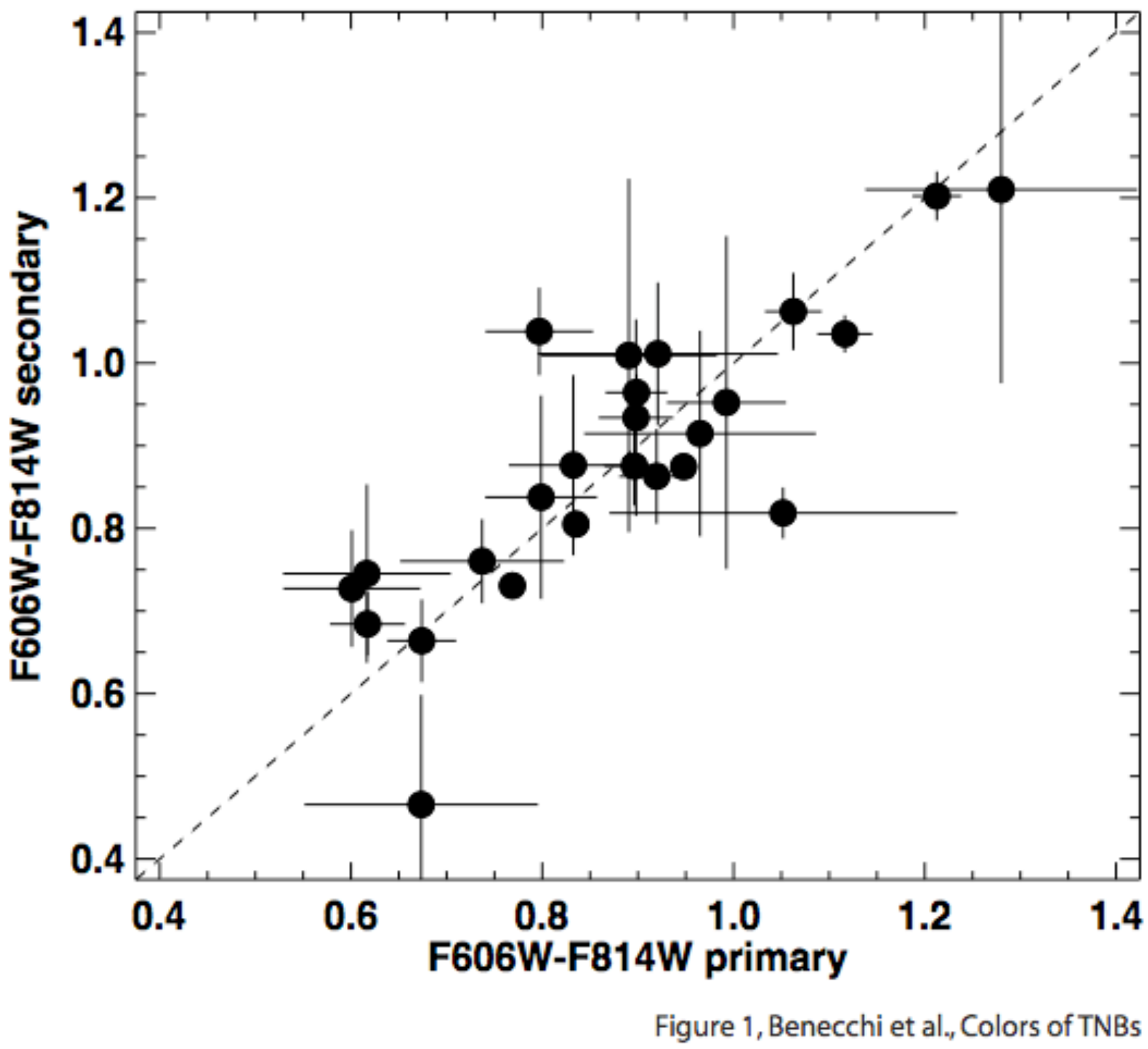

Figure 1. The secondary vs. primary colors for each binary object are plotted. The dashed line demarks a slope of 1, indicating components of identical color. A Spearman rank test shows the primary and secondary colors correlated at the 99.99991% level (5 sigma for a normal distribution). A Pearson product-moment calculation yields a similar result.

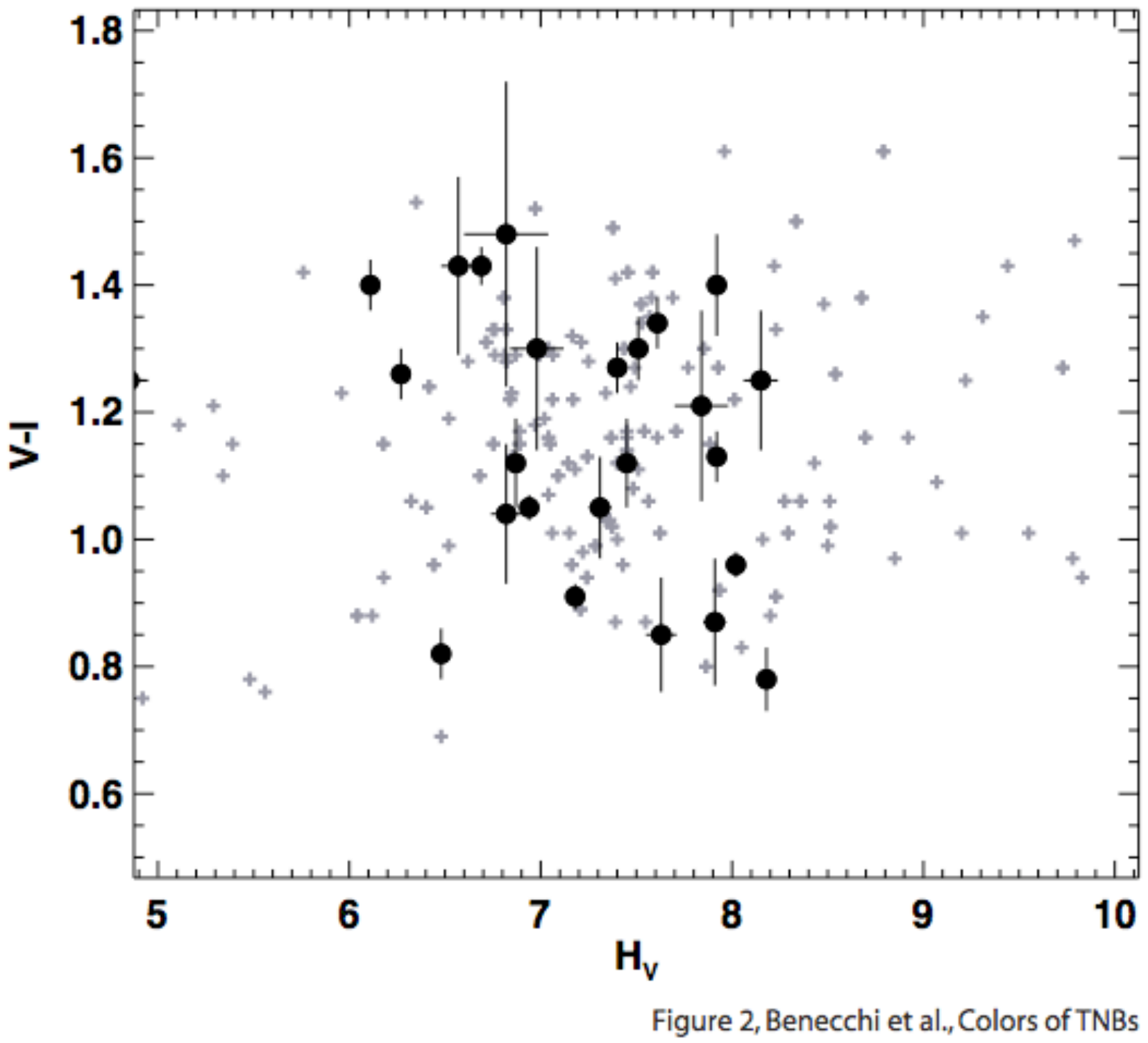

Figure 2. V-I color vs. absolute magnitude is plotted. Only the colors of the primaries (circles) from this work are plotted for clarity. The binary colors occupy the same range as the colors of assumed single TNOs (grey crosses) taken from the MBOSS (Hainaut & Delsanti 2002) and HST (Stephens 2007) databases. Uncertainties on MBOSS colors are not shown for clarity, but are typically about ±0.05 magnitudes.

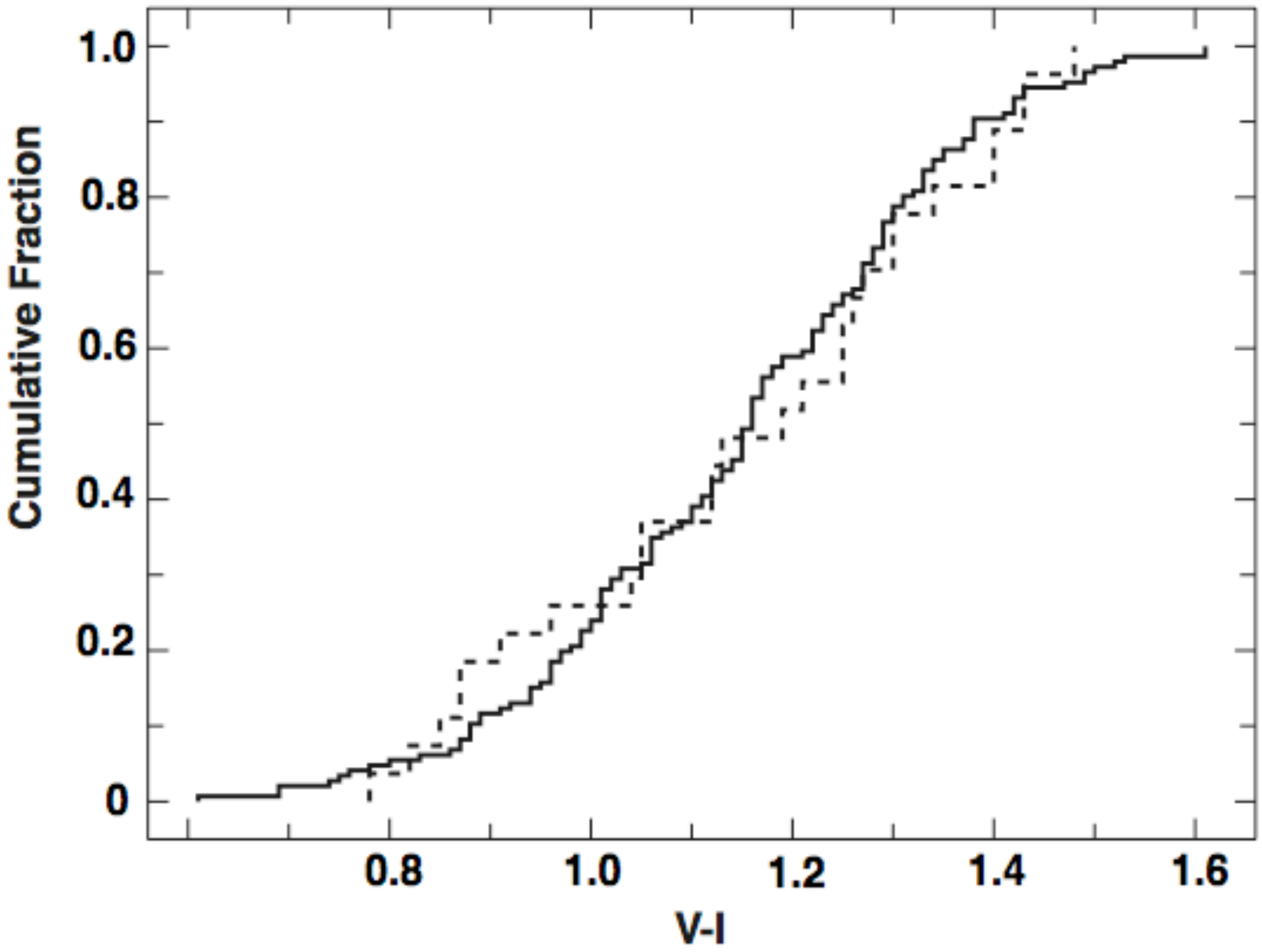

Figure 3. K-S test for all MBOSS (Hainaut & Delsanti 2002) and HST (Stephens 2007) colors (V-I) compared to our HST binary color sample. Without selecting the objects with respect to their dynamical class we find a correlation probability of 96%. If we match the samples dynamically (same percentage of Cold Classical and non-cold-Classical objects in both the MBOSS+HST and TNB sample) we find that the correlation probability ranges from 60-96%. Clearly the TNB components span the same color range as single TNOs implying that whatever colors these objects affects binaries similarly.

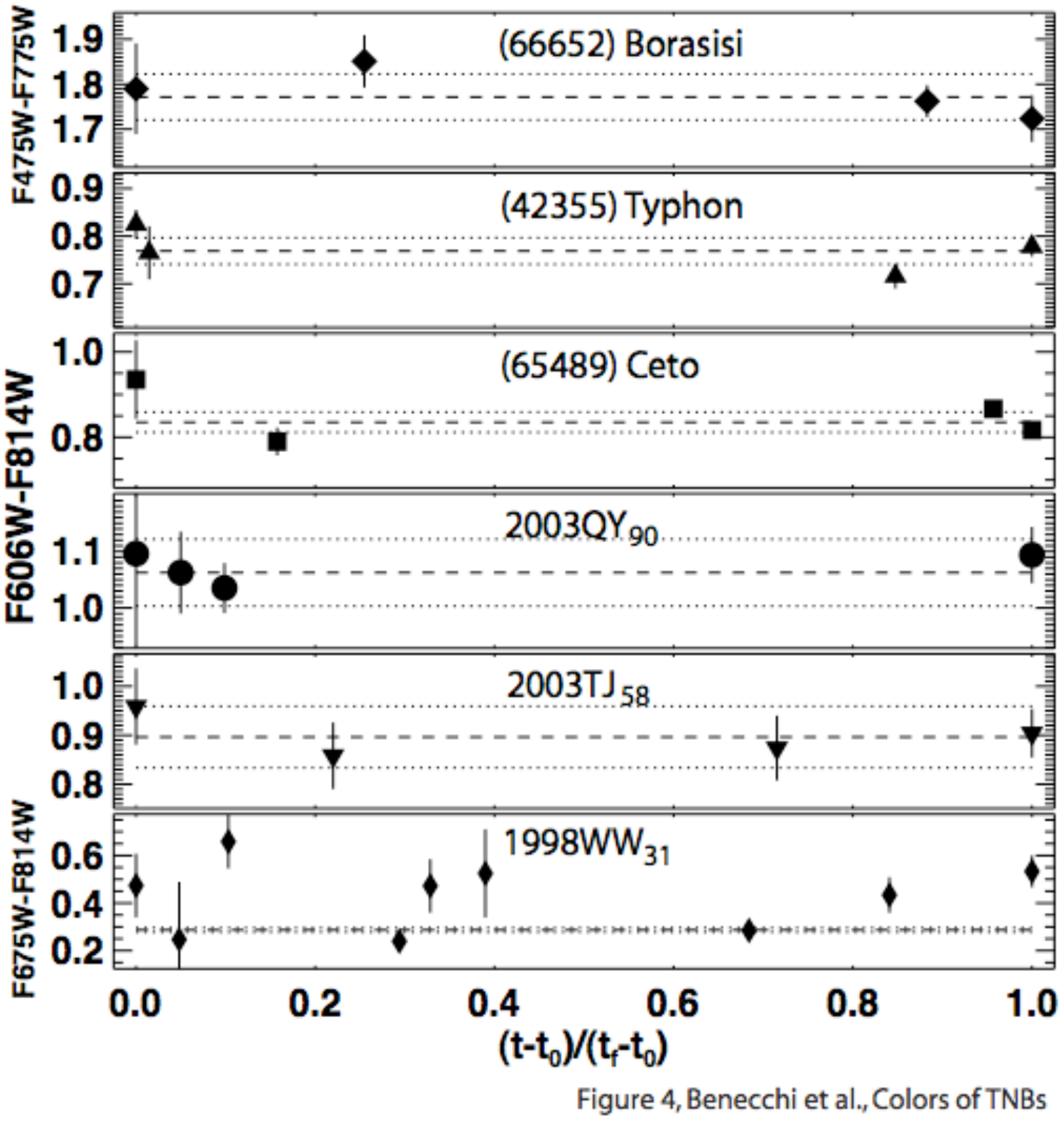

Figure 4. Primary colors obtained at multiple epochs, t, are plotted as a function of days from the first observation $t_0$, normalized by the duration of observation for each object, $t_f$ [(66652) Borasisi $t_f$=100.689 days, (42355) Typhon $t_f$=306.380 days, (65489) Ceto $t_f$=24.757 days, 2003 $QY_{90}$ $t_f$=419.776 days, 2003 $TJ_{58}$ $t_f$=78.249 days, and 1998WW31 $t_f$=226.333 days]. A dashed line is drawn at the weighted mean color for each object with a dotted line at the two sigma level above and below the average. (42355) Typhon, (65489) Ceto and 1998 $WW_{31}$ each have at least one color measurement that deviates from the mean color by 2–sigma (the average individual error bar is 0.03-0.05 magnitudes).